\documentclass[twocolumn,showpacs,amsmath,aps]{revtex4}
\usepackage{txfonts}
\usepackage{graphicx,amssymb,mathrsfs,bm,color,times}

\newcommand{\nl}{\nonumber \\}
\newcommand{\be}{\begin{equation}}
\newcommand{\ee}{\end{equation}}
\newcommand{\bea}{\begin{eqnarray}}
\newcommand{\eea}{\end{eqnarray}}
\newcommand{\bsube}{\begin{subequations}}
\newcommand{\esube}{\end{subequations}}

\newcommand{\Eq}[1]{Eq.\,(\ref{#1})}

\newcommand{\la}{\langle}
\newcommand{\ra}{\rangle}

\newcommand{\dg}{\dagger}
\makeatletter %生成罗马数字
    
    \newcommand{\Rmnum}[1]{\expandafter\@slowromancap\romannumeral #1@}
\makeatother

\begin{document}
\title { Generating and stabilizing the GHZ state in circuit QED:
         Joint measurement, Zeno effect and feedback   }

\author{Wei Feng, Peiyue Wang, Xinmei Ding, Luting Xu and Xin-Qi Li }
        %and Xin-Qi Li \footnote{E-mail: lixinqi@bnu.edu.cn}}
        %\footnote{E-mail:guolingzhen@mail.bnu.edu.cn}
\email{lixinqi@bnu.edu.cn}
\affiliation{Department of Physics, Beijing Normal University,
Beijing 100875, China}

\date{\today}

\begin{abstract}
In solid-state circuit QED system, we extend the previous study
of generating and stabilizing two-qubit Bell state
[Phys. Rev. A {\bf 82}, 032335 (2010)], to three-qubit GHZ state.
In dispersive regime,
we employ the homodyne {\it joint readout} for multiple qubits
to infer the state for further processing,
and in particular use it to stabilize the state directly
by means of an {\it alternate-flip-interrupted} Zeno (AFIZ) scheme.
Moreover, the state-of-the-art feedback action based on
the filtered current enables not only a deterministic generation
of the pre-GHZ state in the initial stage, but also a fast recovery
from occasional error in the later stabilization process.
We show that the proposed scheme can maintain the state
with high fidelity if the efficient quantum measurement
and rapid single-qubit rotations are available.
\end{abstract}

%%\pacs{03.67.-a,32.80.Qk,42.50.Lc,42.50.Pq}

\maketitle

%% ************************************************************************
%% ************************************************************************

%%\section{Introduction}
%{\bf Introduction.}---
%%
{\flushleft
Quantum entanglement is novel and useful because it exhibits correlations that
have no classical analog, and is one of the key ingredients for quantum
technology applications such as quantum teleportation, quantum cryptography,
quantum dense coding, and quantum computation \cite{Nie00}.
Conventional approach to generate quantum entanglement is via unitary
two-qubit gates, using the necessary qubit-qubit interactions.
Interestingly, instead of employing entangling gates, one can exploit
quantum measurement as an alternative means to achieve similar goal.
In both cavity and circuit QED systems, promising ideas along this route
were proposed to {\it probabilistically} create entangled states
by means of the homodyne measurement alone \cite{Mab04,Mab08,Gam08,Gam09}.  }

The circuit QED system \cite{Bla04,Sch04,Mooij04},
a solid-state analog of the conventional quantum optics cavity QED,
is a promising solid-state quantum computing architecture.
This architecture couples superconducting electronic circuit
elements, which serve as the qubits, to harmonic oscillator modes of
a microwave resonator, which serve as a ``quantum bus" that mediates
inter-qubit coupling and facilitates quantum measurement for the qubit state.
Moreover, quantum measurement in this system can be carried out
by operating in the dispersive limit, i.e., with a detuning between
the resonator and the qubit much larger than their coupling strength.
In this limit, the qubit-resonator interaction induces
a qubit-state-dependent shift on the resonator's frequency.
Then, by measuring the resonator output voltage with a homodyne scheme,
information about the qubit state is obtained.

With these advantages,
the schemes proposed in Refs.\ \cite{Gam08,Gam09}
for using measurement to create, respectively, two- and three-qubit
entangled states in the circuit QED system, are attractive.
However, in addition to the drawback of being {\it probabilistic},
the {\it measurement-only} approach cannot stabilize the generated state.
To resolve this problem, the technique of quantum feedback control
may emerge as a possible route \cite{WM10}.
In a recent work \cite{Li10}, a feedback scheme was analyzed for the
creation and stabilization of the two-qubit Bell states.
Owing to using the dispersive joint-readout of multiple qubits
and performing proper feedback,
the scheme leads to results superior to some previous ones.
For instance, it enhances the concurrence to values higher than 0.9,
by noting the 0.31 obtained in Ref.\ \cite{WJ05};
and also it avoids the experimental difficulty in the jump-based
feedback \cite{Car0708},
or the complexity for a state-estimation feedback
\cite{Mab04,Doh99,Ral04-a}.

In this work, we extend the study of generating and stabilizing the
two-qubit Bell state in Ref.\ \cite{Li10} to the three-qubit GHZ state.
To our knowledge, unlike the feedback control of two-qubit Bell state
\cite{Li10,WJ05,Car0708,Ral04-b,Ral08},
schemes for stabilizing three-qubit GHZ state are not well investigated.
In Ref.\ \cite{Mab04} elegant analysis is carried out
for preparing the Dicke state of an ensemble of atoms (qubits) in cavity,
by means of projective measurement in dispersive limit,
and as well by using feedback to make the scheme {\it deterministic}.
However, the ability of stabilizing the generated Dicke state
against decoherence is not demonstrated.
In the present work, we plan to exploit the advantages of the
homodyne {\it joint readout} in dispersive regime for multiple qubits,
to infer the state for further processing,
and in particular to stabilize the target state directly by means of
an {\it alternate-flip-interrupted} Zeno (AFIZ) scheme.
Also, the state-of-the-art feedback action properly
designed according to the measurement current
enables both a {\it deterministic} generation
of the pre-GHZ state in the initial stage, and
a fast recovery from occasional error
in the later stabilization process.

Before proceeding to the details of the proposed scheme,
we first briefly outline the control efficiency.
For the deterministic generation of the pre-GHZ state,
keeping track of the joint measurement information
together with simple individual qubit rotations in our scheme
will either lead to a direct subsequent success of target state generation
with probability $1/2$, being higher than $1/4$ given by the naive rerunning
scheme which additionally needs the difficult ``data clearing" procedures,
or avoid to clear the wrong state before rerunning the generation scheme.
More importantly, the AFIZ stabilization protocol, based only on
an alternate but regular qubit flips and continuous measurement,
can maintain very high fidelity (higher than ninety percent)
for considerably long time (much longer than the single-qubit
decoherence time).
In principle, if the continuous measurement can approach the effect
of fast repeated strong projective measurement, i.e., the ideal Zeno effect,
the AFIZ scheme can stabilize the pre-GHZ state for arbitrarily long time.
Meanwhile, another remarkable advantage of the AFIZ scheme is that it
greatly simplifies the unitary manipulations on qubits,
compared to either the state-based or current-based continuous feedback.
Moreover, associated with the AFIZ scheme, an auxiliary alarm
to the failure of AFIZ control, which may occur {\it occasionally}
owing to the finite strength of the measurement,
can restart the fast deterministic generation
of the target state by simply using two-qubit flips,
one qubit ($\pi/2$) rotation, and two projective measurements.

%%% ================================================================
%%% ================================================================

\begin{figure}
 \begin{center}
 \includegraphics[scale=0.5]{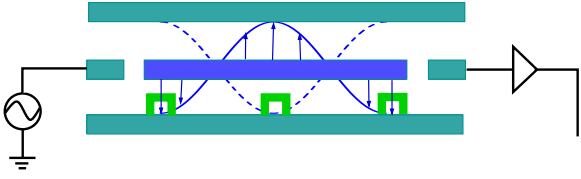}
\caption{\label{}
Schematic diagram of the circuit QED with three qubits,
together with a microwave transmission measurement
in dispersive limit.
The Cooper-pair box qubits are fabricated inside a superconducting
transmission-line resonator and are capacitively coupled to the voltage
standing wave. }
\end{center}
\end{figure}

\vspace{0.2cm}
%%\section{Model and Formalism}
{\bf Model and Formalism.}---
We consider a specific circuit QED system consisting
of three superconducting qubits coupled
to the fundamental mode of a microwave resonator cavity.
The qubits, the resonator cavity and their mutual coupling are
well described by the Jaynes-Cummings Hamiltonian \cite{Bla04,JCM}:
\begin{equation}\label{JCH}
H=\omega_r a^\dagger a +\mathcal{E}(a^\dagger + a)
+\sum^3_{j=1}\left[\frac{\Omega_{j}}{2} \sigma_j^z
+ g_j(\sigma_j^-a^\dagger+\sigma_j^+a)\right].
\end{equation}
The operators $\sigma^-_j(\sigma^+_j)$ and $a(a^\dagger)$ are,
respectively, the lowering (raising) operators for the $j$th qubit
and the resonator cavity photon
(hereafter we would name it as {\it cavity photon} for simplicity).
$\omega_r$ is the frequency of the cavity photon,
and $\Omega_{j}$ and $g_j$ are the $j$th qubit transition energy
and its coupling strength to the cavity photon.
In this work we consider a three-qubit setup as shown schematically
in Fig.\ 1, which can result in the required sign,
$sgn(g_1)=sgn(g_2)=-sgn(g_3)$.
The $\mathcal{E}$-term in \Eq{JCH} stands for a microwave driving
to the resonator cavity that is employed for the task of measurement.
Explicitly, $\mathcal{E}=\epsilon e^{-i\omega_m t}+{\rm c.c}$,
where the driving frequency can differ from the cavity photon frequency,
i.e., $\Delta_r\equiv\omega_r-\omega_m\neq 0$.
Moreover, we will focus on a dispersive limit measurement
\cite{Bla04,Sch04,Mooij04}, i.e., with the energy detuning
$\Delta_j=\omega_r-\Omega_j$ much larger than $g_j$.
In this limit, the canonical transformation,
$H_{\rm eff}\simeq U^{\dg}HU$, where
$U=\exp[\sum_j\lambda_j(a\sigma_j^+ -a^{\dg}\sigma_j^-)]$
with $\lambda_j=g_j/\Delta_j$, yields
(in a {\it joint rotating frame} with the driving frequency $\omega_m$
with respect to both the cavity photon and each individual qubit)
\bea\label{H_eff}
H_{\rm eff}&\simeq& \Delta_ra^{\dg}a
  +\left(\epsilon^\ast a+\epsilon a^{\dag}\right)
  + \sum^3_{j=1}(\omega_j+\chi_j)\frac{\sigma^z_j}{2}
  + \sum^3_{j=1}\chi_j a^{\dg}a\sigma^z_j   \nl
%  + \sum_{j>k} J_{jk}(\sigma^+_j\sigma^-_k+\sigma^-_k\sigma^+_j) ,
\eea
where $\omega_j=\Omega_j-\omega_m$, and $\chi_j=g_j^2/\Delta_j$.
Here we have neglected the virtual cavity-photon-mediated effective coupling
between qubits, as is appropriate for sufficient detuning between qubits
\cite{Gam09}.

In the circuit-QED system, the measurement is typically performed
via a homodyne detection of the transmitted microwave photons.
The photon's leakage from the resonator cavity
is described by a Lindblad term $\kappa {\cal D}[a]\rho$
in the master equation, where $\kappa$ is the leakage rate
and the Lindblad superoperator acting on the
reduced density matrix $\rho$ is defined by
${\cal D}[a]\rho=a\rho a^{\dg}-\frac{1}{2}\{a^{\dg}a,\rho\}$.
However, conditioned on the output homodyne current, i.e.,
$I_{\rm hom}(t)=\kappa\la a+a^{\dg} \ra_c(t)+\sqrt{\kappa}\xi(t)$,
there will be an additional unravelling term in the conditional
master equation, $\mathcal{H}[a]\rho_c \xi(t)$.
Here, $\la (\cdots) \ra_c(t)\equiv {\rm Tr}[(\cdots)\rho_c(t)]$
with $\rho_c(t)$ the conditional density matrix,
and $\mathcal{H}[a]\rho_c \equiv a\rho_c+\rho_c a^{\dg}
-\mathrm{Tr}[(a+a^{\dg})\rho_c]\rho_c$.
And, the Gaussian white noise $\xi(t)$, which has the ensemble average
properties $E[\xi(t)]=0$ and $E[\xi(t)\xi(t')]=\delta(t-t')$,
stems from the quantum-jump related stochastic nature.

In this work, we assume a {\it strongly damped} resonator cavity,
which enables to adiabatically eliminate the cavity photon
degree of freedom \cite{Gam08,WMJW9394}.
Qualitatively, observing the effective coupling
$\sum^3_{j=1}\chi_j a^{\dg}a\sigma^z_j$ in \Eq{H_eff},
we can understand that the fluctuation of the photon number
will cause a pure dephasing backaction onto the qubits,
with a joint dephasing operator $J_z=\sum^3_{j=1}\delta_j \sigma^z_j$,
where $\delta_j=\chi_j/\bar{\chi}$ and $\bar{\chi}=\sum^N_{j=1}\chi_j/N$.
Thus, we can expect the following results after adiabatic elimination
of the photon's degree of freedom:
{\it (i)} - the dephasing term
$\sim \mathcal{D}[J_z]\rho_c$;
{\it (ii)} - the unravelling term $\sim\mathcal{H}[J_z]\rho_c \xi(t)$;
and {\it (iii)}
- the homodyne current $ I_{\rm hom}(t)\sim \la J_z\ra_c(t)+\xi(t)$.
Indeed, following the standard procedures of adiabatic elimination
\cite{Gam08,WMJW9394}, an {\it effective} quantum trajectory equation (QTE)
involving only the degrees of freedom of qubits can be obtained as
\begin{eqnarray}\label{QTE}
 \dot{\rho}_c = \mathcal{L}\rho_c
  +\frac{\Gamma_{d}}{2}\mathcal{D}[J_z]\rho_c
  +\frac{\sqrt{\Gamma_m}}{2}\mathcal{H}[J_z]\rho_c\xi(t),
\end{eqnarray}
in which the Liouvillian is defined as
\begin{eqnarray}\label{L}
&&\mathcal{L}\rho_c = -i\left[\sum_{j}\frac{\omega_{j}
   +\chi_j}{2}\sigma_j^z+\bar{\chi}|\alpha|^2\sum_{j}
   \delta_j\sigma_j^z, ~~\rho_c \right]  \nl
 && ~~~~~~
 + \sum_{j} (\gamma_{j}+\gamma_{pj})\mathcal{D}[\sigma_j^-]\rho_c
    +\sum_{j}\frac{\gamma_{\phi j}}{2}\mathcal{D}[\sigma_j^z]\rho_c  \nonumber.
\end{eqnarray}
Here we have also assumed a resonant driving, i.e., $\Delta_r=0$.
In \Eq{QTE}, $\gamma_j$ and $\gamma_{\phi j}$ are the relaxation and dephasing
rates caused by the surrounding environments. Since the external dephasing can
be strongly suppressed by proper design of the superconducting qubits,
we will thus neglect it in our following simulations.
The $\gamma_{pj}$ terms, with $\gamma_{pj}=\kappa\lambda_j^2$,
stem from the so-called Purcell effect, describing an {\it indirect}
qubit decoherence induced by the damping of the cavity photons.
Since we assumed mutually distinct frequencies of qubits,
we can treat the Purcell effect under secular approximation,
i.e., neglecting the cross terms in
$\mathcal{D}[\sum_j g_j\sigma_j^-]\rho_c$.
(The cross terms characterize {\it interference} between the radiation
from different qubits, and become important only if the qubit
frequencies are sufficiently close.)
With these considerations, the Purcell-effect-induced and
other environment-caused decoherences can be equally treated.
Therefore, we combine them by $\gamma=\gamma_j+\gamma_{pj}$.
Moreover, the measurement-backaction induced dephasing rate
$\Gamma_d=8|\alpha|^2 \bar{\chi}^2/\kappa$, with $\alpha=-2i\epsilon/\kappa$.
And finally, the information-gain rate $\Gamma_m$ in \Eq{QTE} is in general
related to the backaction dephasing rate in terms of the quantum efficiency,
$\eta=\Gamma_m/(2\Gamma_{d})$.

%% ================================================

\vspace{0.2cm}

{\bf Filtering the Output Current.}---
After adiabatic elimination of the cavity degree of
freedom, we obtain an effective measurement operator
$J_z=\sum^3_{j=1}\delta_j \sigma^z_j$
which has a number of discrete eigenvalues.
However, in the practical homodyne measurement, the output record
is the homodyne current which we rewrite as
$dI_{hom}(t)=\sqrt{\Gamma_m}\langle J_z \rangle_c(t) dt+dW(t)$,
where $dW(t)=\xi(t)dt$ is the Wiener increment that has the
statistical properties $E[dW(t)]=0$ and $E[dW(t)dW(s)]=\delta(t-s) dt$.
In this context, we remind that we are actually employing
a series of photons by using their transmission through
the cavity to measure the qubit state.
Accordingly, the stochastic Wiener increment in the homodyne current
just characterizes the very weak (partial) random collapse caused by
the individual measuring photons.
It is the sum of the $\langle J_z \rangle_c(t)$ term plus the
Wiener increment that corresponds to the measurement record
{\it after} the (infinitesimal) partial collapse,
while $\langle J_z \rangle_c(t)$, from its definition
$\langle J_z \rangle_c(t)={\rm Tr}[J_z\rho_c(t)]$,
exposes the information of state {\it before} the infinitesimal collapse.

Of great interest and very usefully,
instead of knowing $\langle J_z \rangle_c(t)$
from the usual ensemble measurement by repeating a large number
of realizations (noting that the quantum mechanical expectation
indicates a statistical or ensemble-average interpretation),
we can approximately obtain it in the context of continuous weak measurement
by averaging the homodyne current over a properly chosen window of time.
This has some similarity to the {\it ergodic} assumption in statistical physics,
where the average along time is assumed to replace the ensemble average.
In practice, we can low-pass filter the homodyne current
over a small time window \cite{Mil05}, and get a
smoothed signal as
$\bar{I}_{\rm hom}(t)=\frac{1}{\mathcal{N}}\int_{t-T}^t
  e^{-\gamma_{\rm ft}(t-\tau)}  dI_{\rm hom}(\tau)$,
where $\gamma_{\rm ft}$ is a low-pass filtering parameter,
and the factor $\mathcal{N}$ normalizes the smoothed signal
to a maximum magnitude of unity.
Numerical test shows that $\bar{I}_{\rm hom}(t)$
indeed coincides with $\la J_z \ra_c(t)$
satisfactorily. Thus, in the numerical simulations of this work, we simply use
$\la J_z \ra_c(t)$ as a state indicator to guide our feedback manipulations.
In particular, after the qubits have experienced sufficient transmission
measurement by a large number of photons and fully collapse onto one
eigenstate of the measurement operator, we can in principle
unambiguously infer it from the filtered current $\langle J_z \rangle_c(t)$.

Finally, we remark that, in regard to feedback design,
$\la J_z\ra_c(t)$ in the homodyne current $dI_{\rm hom}(t)$
is informative about the state and is thus useful,
while the $dW(t)$ term is yet harmful.
In other words, the $dW(t)$ term in the homodyne current
has {\it distinct roles} when using the current to update
the state {\it versus} to perform feedback.
In doing the former, it is necessary; while in doing the latter,
it is useless and should better be erased.
This understanding was demonstrated in the recent work \cite{Li10}.

%% ================================================

\vspace{0.2cm}
{\bf Deterministic Generation of the Pre-GHZ State}.---
Since our control target, say, the GHZ state $|000\rangle+|111\rangle$,
can be easily obtained by a simple flip of single qubit
(i.e. the third one) from $|001\rangle+|110\rangle$
which may be accordingly named as {\it pre-GHZ} state,
we can aim to control this pre-GHZ state instead.
From the QTE formalism outlined above in dispersive regime,
we see that an effective measurement operator reads
$J_z=\sum_{j}\delta_j\sigma_j^z$.
Noting that the consequence of a quantum measurement is to collapse
an arbitrary state onto one of the eigenstates of the measurement operator,
we may design $J_z=\sigma_1^z+\sigma_2^z+2\sigma_3^z$,
which makes the pre-GHZ state be its eigenstate with eigenvalue $J_z=0$.
In practice, this can be realized by setting the dispersive shifts
$\chi_1:\chi_2:\chi_3=1:1:2$.

More specifically, let us start with an initially separable state:
\begin{eqnarray}
|\Psi_i\ra
 &=& \frac{1}{\sqrt{2}}(|0\ra+|1\ra)_1\otimes
 \frac{1}{\sqrt{2}}(|0\ra+|1\ra)_2\otimes\frac{1}{\sqrt{2}}(|0\ra+|1\ra)_3   \nl
 &=&\frac{1}{\sqrt{8}} [~|000\rangle+|111\rangle+(|010\rangle+|100\rangle)   \nl
 & & + ~(|011\rangle+|101\rangle)+(|001\rangle+|110\rangle)~]   .
\end{eqnarray}
Performing the above designed homodyne measurement,
in an individual single realization,
would collapse $|\Psi_i\ra$ {\it stochastically} onto
one of the eigenstates of $J_z$.
According to the principle of quantum projective measurement,
$|\Psi_i\ra$ would collapse
onto the pre-GHZ state $|001\rangle+|110\rangle$ with probability $1/4$,
as a result of getting record $J_z=0$.
However, there are probabilities to get other results.
That is, the state would collapse onto $|011\rangle+|101\rangle$
or $|100\rangle+|010\rangle$ with probability $1/4$,
depending on the result to be $J_z=2$ or $-2$.
It may also collapse onto $|111\rangle$ or $|000\rangle$
with probability $1/8$ if one gets $J_z =4$ or $-4$.

What we described above is in fact a {\it measurement alone} scheme
to generate the pre-GHZ state {\it stochastically}.
Below we show that proper current-based feedback
manipulations can make the scheme {\it deterministic}.
If, unfortunately, we did not get the result $J_z=0$, we may
adopt the following distinct procedures based on the specific
measurement results obtained.
{\it (i)}
- If the result is $J_z=2$, which indicates the state
$|011\rangle+|101\rangle$ projected out,
we perform a $\sigma_x$-flip on the first qubit
and a $\pi/2-\sigma_y$ rotation on the third qubit.
Noting that $|011\rangle+|101\rangle$ can be rewritten as
$(|01\rangle+|10\rangle)\otimes|1\rangle $,
it is clear that the above rotations will transform it to
$|000\rangle+|111\rangle+|001\rangle+|110\rangle$,
which then has a new probability of $1/2$ in the successive measurement
to be collapsed onto the pre-GHZ state $|001\rangle+|110\rangle$.
{\it (ii)}
- Similarly, for the result $J_z=-2$, a $\sigma_x$-flip on the first
qubit and a $3\pi/2-\sigma_y$ rotation on the third one
can be performed to achieve the same goal as described in {\it (i)}.
{\it (iii)}
- If the measurement result is $J_z =4$ or $-4$,
which indicates the state $|111\rangle$ or $|000\rangle$ obtained,
we then apply a $\pi/2 - \sigma_y $
or a $3\pi/2 - \sigma_y $ rotation on each qubit,
making the state return back to the initial one
$(|0\ra+|1\ra)_1\otimes(|0\ra+|1\ra)_2\otimes(|0\ra+|1\ra)_3$
that allows to rerun the generating procedures.

%% =================================================================

\begin{figure}
%%\center
\begin{center}
\includegraphics[width=0.55\textwidth,bb=0 0 400 300]{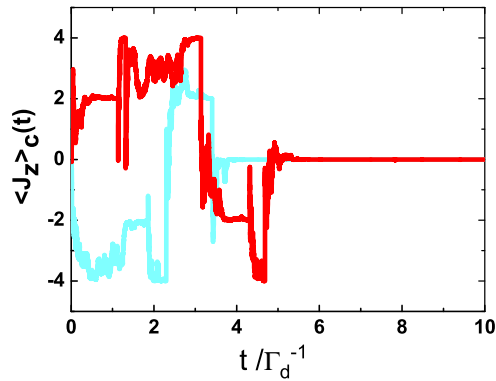}
\caption{(color online)
Two representative quantum trajectories showing
the deterministic generation of the pre-GHZ state.  }
\end{center}
\end{figure}

In Fig.\ 2 we show two representative quantum trajectories
which are {\it deterministically} guided to the pre-GHZ state.
The quantity $\la J_z\ra_c(t)$ plotted here is an appropriate
indicator for the finally collapsed state. Here, we remark
that our present scheme of entangled state generation is {\it efficient}.
Generally speaking,
probabilistic scheme of quantum information processing
needs to recycle the process with the same initial state.
This will require a procedure of clearing the {\it unwanted data}
(qubit states), which is, unfortunately, not an easy job
in quantum system compared to its classical counterpart.
On the contrary, in addition to that we do not require ``data clearing",
our scheme in case {\it (ii)} only needs two single-bit rotations
and the subsequent success probability of projection (i.e., 1/2)
is higher than the recycling scheme which has a success probability $1/4$.
Even in the worst case {\it (iii)}, the manipulation of qubit
rotations is identical to rerunning the generation procedure,
but the process of ``data-clearing" is avoided.

%% =================================================================
%\clearpage

\vspace{0.2cm}
%\leftline{\bf $\S\S$  Stabilization: measurement only and feedback involved }
%\vspace{0.2cm}
{\bf Stabilization Using Flip-Interrupted Zeno Projection and Feedback}.---
In the above, we discuss an efficient \emph{deterministic} scheme
to generate the pre-GHZ state. However, under the unavoidable influence
of the surrounding environment, this state will degrade
if we do not provide proper active protections.
Below we first propose an {\it alternate-flip-interrupted} Zeno (AFIZ)
stabilization scheme
to enhance considerably the lifetime of the pre-GHZ state,
then design additional manipulation to prevent the state from degradation.
Actually, the pre-GHZ state is an eigenstate of the measurement operator $J_z$.
Then, if one performs a continuous observation (measurement) on it,
the quantum Zeno effect would attempt to freeze the state.
Using Zeno effect to stabilize quantum state is an interesting topic
in quantum physics and particularly in quantum computing \cite{Vai96,Bra97}.
However, as we will see shortly,
generalization of the Zeno protection from single
to multiple qubits will suffer more complexities.
As a major contribution of the present work, we will see that the proposed
AFIZ stabilization scheme can greatly improve the control quality.

\begin{figure}
 \center
 \includegraphics[scale=0.45]{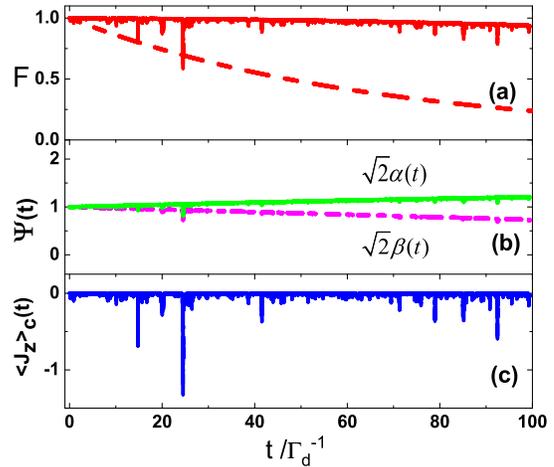}
 \caption{(color online)
(a)
State fidelity under the conventional quantum Zeno
(not the AFIZ) stabilization for the pre-GHZ state,
showing the result (solid line) much better
than the uncontrolled one (dashed line).
(b)
Detailed inspection for the Zeno pulled-back state in (a),
$|\Psi(t)\ra = \alpha(t) |001\ra  +  \beta(t) |110\ra $,
showing a gradual deviation from the target state
$|\Psi_T\ra =(|001\ra  +  |110\ra)/\sqrt{2} $.
(c)
Unconscious output current for the changing state $|\Psi(t)\ra$.
Single qubit decoherence rate: $\gamma=0.01\Gamma_d$.   }
\end{figure}

In Fig.\ 3(a) we plot the state fidelity under Zeno protection
against the one in the absence of such protection,
where the Zeno effect is {\it automatically} realized via the
continuous $J_z$-type measurement as discussed above.
We see that, provided the measurement strength is much stronger
than the decoherence rate, the effect of Zeno stabilization is obvious.
Regarding the underlying mechanism of Zeno stabilization for
multiple qubit state, it is analogous to that for single qubit.
That is, while the environment is causing the state to
leave away from the target state to other unwanted states,
the relatively strong continuous measurement
is at the same time pulling it back.
Since our target state is a superposition of $|001\ra$ and $|110\ra$,
it should, however, be more fragile than the {\it one-component} state
in regard to the Zeno stabilization.

We may understand this point better as follows.
Owing to coupling with environment, the qubits would experience
an entangling evolution with the environment, for an infinitesimal
interval of time. Then, the measurement projects the qubit state back.
Since each individual qubit couples to the environment independently,
this pull-back action via measurement from entangling with the
environment cannot guarantee the {\it reorganized} components $|001\ra$ and
$|110\ra$ with exactly {\it unchanged} superposition weights.
And, unfortunately, the changed superposition amplitudes cannot be
distinguished by the measurement, since the arbitrary superposition
of $|001\ra$ and $|110\ra$ is the eigenstate of the measurement operator
and will result in the same output current.
In Fig.\ 3(b) and (c) we show, respectively, the gradual change of
the superposition amplitudes away from the initial value $1/\sqrt{2}$,
and the corresponding output current.

The Zeno effect induced pull-back action, in terms of quantum measurement
language, corresponds to a null-result record of spontaneous emission of
the qubits. Conditioned on the null result of spontaneous emission,
an effective Hamiltonian governing the qubits state evolution reads
$\tilde{H}_{\rm qu}=H_{\rm qu}-i \frac{\gamma}{2}\sum^3_{j=1} \sigma^+_j\sigma_j$,
where $H_{\rm qu}$ stands for the qubits Hamiltonian contained in \Eq{H_eff}.
With this effective Hamiltonian acting on
the target pre-GHZ state $|\Psi_T\ra=(|001\ra + |110\ra)/\sqrt{2}$,
or more generally on $|\Psi\ra=\alpha|001\ra + \beta|110\ra$,
leads to an effective evolution as
$|\Psi(t)\ra=(\alpha e^{-\gamma t/2} |001\ra
+ \beta e^{-\gamma t} |110\ra)/||\cdot||$,
where $||\cdot||$ denotes the normalization factor.
We have checked that this effective evolution is in perfect agreement
with the results from numerical simulation, e.g., the one shown in Fig.\ 3(b).
%%%%
%%%%
Then, favorably, we get an insight that the change
of the superposition amplitudes is owing to the {\it unbalance}
of qubit state ``0" and ``1" in the components $|001\ra$ and $|110\ra$.
Based on this observation, quite simply, if we flip simultaneously
each of all the three qubits after the Zeno stabilization
continuing for a time period $\tau$,
the pre-GHZ state will be restored from the flipped state
after an equal time interval  $\tau$.
In practice, the time interval $\tau$ can be chosen as a few $\Gamma^{-1}_d$,
since in such a short timescale the variations of the state amplitudes
$\alpha$ and $\beta$ are negligibly small, provided $\gamma<<\Gamma_d$.
We can expect and will demonstrate in the following
that this {\it alternate evolution}, which is named in this work
as an {\it alternate-flip-interrupted} Zeno (AFIZ) scheme,
is capable of stabilizing the pre-GHZ state very efficiently.
As an interesting remark, this AFIZ scheme has certain similarity
to the {\it refocus} technique in the echo physics
in quantum optics or nuclear magnetic resonance,
where similar flip is manipulated to cause an inverse evolution
towards the initial state, i.e., to produce an echo.

%
%To improve the stabilization quality of the pre-GHZ state,
%in addition to the quantum Zeno effect discussed above,
%we propose further the following active feedback procedures.

Unfortunately, to implement the above AFIZ protection in practice,
there should exist very small but nonzero probabilities to collapse
the qubits state to $|000\rangle$ and a mixture of
$|100\rangle$ and $|010\rangle$,
owing to finite strength of the measurement.
{\it (i)} -
For the result of $|000\rangle$, the output current would trigger
a feedback action as described in the deterministic generation scheme,
which will force the state rapidly back to the
pre-GHZ state under the guided efficient projection of measurement.
{\it (ii)} -
For the mixture of $|100\rangle$ and $|010\rangle$,
the output current will also trigger a feedback action
as described in the deterministic generation scheme.
As a result, besides being projected to $|000\rangle$ and $|111\rangle$
which will be further guided to the pre-GHZ state,
a mixed state with $|001\rangle$ and $|110\rangle$
will be filtered out by the measurement.
Very unfortunately, this mixed state is {\it not} the pre-GHZ state,
but with the same zero output current. To eliminate this error,
one can perform a flip action on the third qubit.
Then, a mixed state with  $|000\rangle$ and $|111\rangle$
is formed and the rapid deterministic generation procedures
will be triggered.

%
%In any of such cases happening, the output current would be nonzero.
%Then, using the current as an indicator for the collapsed state,
%a feedback action is to be triggered as described
%in the {\it deterministic} generation scheme.
%
%

\begin{figure}
 \center
 \includegraphics[scale=0.42]{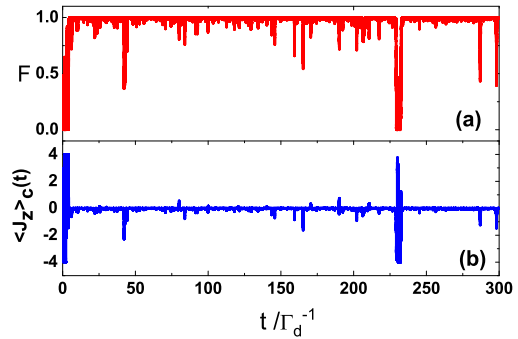}
 \caption{(color online)
(a)
Fidelity of the pre-GHZ state
under the combined AFIZ-plus-feedback stabilization.
(b)
The corresponding output current.
Single qubit decoherence rate: $\gamma=0.01\Gamma_d$. }
\end{figure}

In Fig.\ 4 we show the numerical result of stabilizing the pre-GHZ state,
based on the measurement and feedback schemes described above.
The stabilization dynamics is illustrated by both the state fidelity
and the output current.
We see that at most times the current is zero, only interrupted
occasionally by jumps between $0$, $\pm 2$ and $\pm 4$.
The flat current indicates the stage of Zeno stabilization,
during which the quality of the pre-GHZ state is maintained
at desirable high level, with a state fidelity larger than 0.9.

\begin{figure}
 \center
 \includegraphics[scale=0.5]{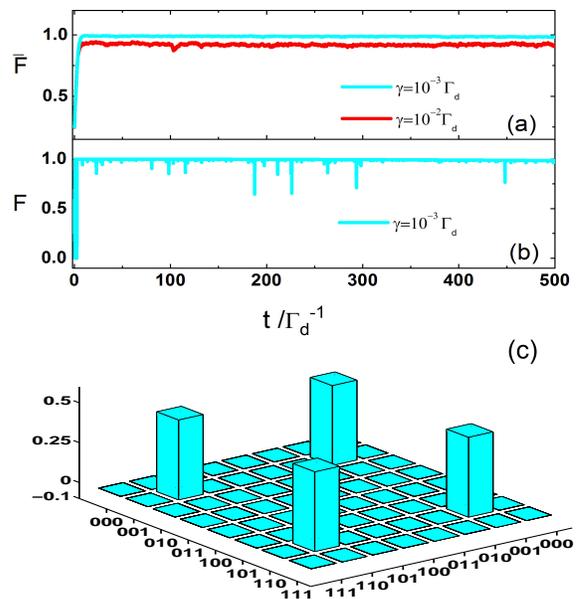}
 \caption{(color online)
(a)
Average fidelity of the pre-GHZ state over 1000 quantum trajectories.
(b)
Fidelity of an individual realization with $\gamma=0.001\Gamma_d$,
showing perfect control result under this even weaker decoherence
when compared to $\gamma=0.01\Gamma_d$ in Fig.\ 4.
(c)
The full state density matrix at a specific time in (b). }
\end{figure}

The state fidelity of single quantum trajectory has certain
stochasticity, owing to the measurement induced quantum jumps.
We may follow the conventional way to employ the fidelity
of the ensemble average state as a reliable figure of merit
to characterize the control quality.
Figure 5(a) shows the results of ensemble average fidelity,
for $\gamma=10^{-2}\Gamma_d$ and $10^{-3}\Gamma_d$, respectively.
We notice that, even for $\gamma=10^{-2}\Gamma_d$,
which in most cases such as the one- or two-qubit feedback control
is taken as a tolerable error rate,
the average fidelity can be higher than 0.9,
while for smaller error rate such as $\gamma=10^{-3}\Gamma_d$
the average fidelity can reach nearly unity
and the individual quantum trajectory also shows perfect control result
as illustrated in Fig.\ 5(b).
In addition, in Fig.\ 5(c) we present the full density matrix
for a representative state during the Zeno stabilization stage,
which clearly reveals the quality of the protected pre-GHZ state.

%% =================================================================
\vspace{0.2cm}
{\bf Concluding Remarks}.---
Finally, we make a number of remarks before summarizing the work.
{\it (i)}
- Some approximations are involved, for instance,
the rotating-wave approximation contained
in the Jaynes-Cummings model Eq.\ (1), the effective
Hamiltonian Eq.\ (2) in a dispersive regime, and the
adiabatic elimination of cavity photons leading to Eq.\ (3).
In the recent work by Liu {\it et al} \cite{Li10},
all these approximations are properly justified.
{\it (ii)}
- The main problem of doing feedback in circuit QED is the lack of
efficient homodyne detection. Currently, the way to perform homodyne
and heterodyne detection is to first amplify the signal before
mixing it on a nonlinear circuit element of some kind.
As a consequence, the extra noise added by the amplifier will
reduce the quantum efficiency and prohibit quantum limited
feedback. It seems that this situation is to be changed quickly,
for instance, by developing Josephson parametric amplifiers
which can be realized in superconducting circuits \cite{Teu09}.
{\it (iii)}
- In our numerical simulation, we did not explicitly include the nonunit
quantum efficiency in the homodyne detection of the field.
After adiabatic elimination of the cavity photon degree of freedom,
the nonunit quantum efficiency of homodyne detection will
reduce the effective information-gain rate $\Gamma_m$ in Eq.\ (3).
This implies an emergence of an extra non-unravelling dephasing
term in the quantum trajectory equation.
However, for the present particular study, this term only results
in dephasing among the pre-GHZ state and the others
in the single quantum-trajectory realizations.
In addition to simple intuitive expectation, we have numerically examined
that lowering the quantum efficiency by some acceptable amount does not
obviously change the results.
{\it (iv)}
- Experimental verification of the GHZ state is of great interest
and is analyzed theoretically in the recent work by Bishop
{\it et al} \cite{Gam09}.
In order to observe a violation of the Bell-Mermin inequality,
relatively high signal-to-noise ratio in performing the measurement
is required, which is unfortunately beyond the existing scope of experiment.
But, optimistically, the situation is expected to change in the near future
by fast experimental progresses.

To summarize, we have presented a promising quantum control scheme
for deterministic generation and stabilization of three-qubit GHZ state
in the solid-state circuit QED system.
The scheme largely depends on a joint-readout of multiple qubits
in dispersive regime,
which enables not only to infer the state for further processing,
but also to stabilize the target state directly by means
of an alternate-flip-interrupted Zeno (AFIZ) projection.
The proposed scheme was demonstrated by quantum trajectory simulations
which show satisfactory control effects.

%% +++++++++++++++++++++++++++++++++++++++++++++++++++++++++++++++++++
%% +++++++++++++++++++++++++++++++++++++++++++++++++++++++++++++++++++
%%\clearpage

\vspace{1cm}
{\it Acknowledgements.}---
This work was supported by the NNSF of China under grants
No.\ 101202101 \& 10874176.

%and the Major State Basic Research Project
%under grant No. 2006CB921201.

%% =============================================================

\end{document}